\pdfoutput=1

\documentclass[
  fontsize=12pt,
  fleqn,
  final
]{scrartcl}

\RequirePackage{metadata/manuscript}

\begin{document}


\thispagestyle{empty}
{
\noindent
\Large
\bfseries
Topological symmetry-restored phase of gravity
\bigskip
\bigskip
}

{
\noindent
\bfseries
Guilherme~Sadovski${}^{1}${\normalfont~and}
Rodrigo~F.~Sobreiro${}^{2}$
\bigskip
}

{
\noindent
\footnotesize
${}^{1}$HENU --- Henan University, Institute of Contemporary Mathematics, School of Mathematics and Statistics, Kaifeng, Henan 475004, P. R. China. \\
${}^{2}$UFF --- Universidade Federal Fluminense, Instituto de F{\'{\i}}sica, Campus da Praia Vermelha, Av. Litor{\^{a}}nea s/n, 24210--346, Niter{\'{o}}i, RJ, Brazil.
\bigskip
\bigskip
}

{
\noindent
\footnotesize
\rmfamily
e-mails:~\href{mailto:gsadovski@henu.edu.cn}{gsadovski@henu.edu.cn}
and
\href{mailto:rodrigo\_sobreiro@id.uff.br}{rodrigo\_sobreiro@id.uff.br}.
\bigskip
}

{
\noindent
\bfseries
Abstract.\normalfont{} In this work, we propose a topological quantum field theory phase for four-dimensional gravity. We show it is able to generate, not only General Relativity, but the whole family of Lovelock-Cartan theories of gravity. This is accomplished due to the existence of a topological symmetry which, when explicitly broken via the introduction of a mass scale, releases the local degrees of freedom of gravity. Additionally, we introduce an extended notion of the (anti-){}self-dual Landau gauge conditions to evaluate the Ward identities, counterterms, and prove the quantum stability of the model, to all orders in perturbation theory, using the algebraic renormalization technique.
\bigskip
}

{
\noindent
\rule{\textwidth}{1pt}
\vspace{-4.5ex}
\tableofcontents
\noindent
\rule{\linewidth}{1pt}
}

\null{}


\section{Introduction}%
\label{sec:introduction}

Topological quantum field theories (TQFTs) can be interpreted as exact path integral representations of many classes of topological invariants~\cite{birmingham1991a,blau1993a,cordes1995a}. Most notably, $ n=2 $ topological $ \sigma $-models\footnote{ $ n $ stands for the number of spacetime dimensions. } are related to Gromov-Witten invariants of pseudo-holomorphic curves, $ n=3 $ Chern-Simons (CS) theory to Jones polynomials of knots and links, $ n=4 $ Topological Yang-Mills (TYM) theory to Donaldson polynomials of smooth 4-manifolds, and so on~\cite{witten1988c,witten1989a,witten1988d}.

By their nature, TQFTs are conformal and diffeomorphism invariant. Additionally, they are well-defined perturbative quantum field theory, often presenting a very simple local dynamics, and having a finite number of degrees of freedom~\cite{brandhuber1994a,werneck1993a,sadovski2018b,sadovski2020a,junqueira2021a}. Gauge TQFTs, in particular, tend to be exactly solvable in the perturbative regime: $ n=3 $ CS theory at 1-loop, and $ n=4 $ TYM at tree-level~\cite{blasi1991a,maggiore1992a,piguet1995a,sadovski2017c,sadovski2018a}.

Due to these features, TQFTs represent an appealing framework to express quantum gravity ideas. In low spacetime dimensions, this program has found the most success. In $n=2$, quantum GR is explicitly topological: its dynamics is given by a sum over genera. Moreover, one can construct the gravitational Labastida-Pernici-Witten model, related to Mumford-Morita-Miller classes~\cite{labastida1988b,labastida1988a}. Both approaches deal with invariants of Riemann surfaces, admit random matrix integral representations, and can be associated with a non-perturbative description of non-critical strings~\cite{douglas1990a,dijkgraaf1991a,gross1990a,witten1990a,kontsevich1992a,dijkgraaf2002a}.

In high spacetime dimensions, these fundamental equivalences become foggier. In $ n=3 $, quantum GR is equivalent to CS theory only in the perturbative limit~\cite{achucarro1986a,witten1988b,witten2007a}. The success of random matrices has not being replicated by a well-behaved geometric limit for random tensors~\cite{gurau2024a,carrozza2024a}. And, the association to non-critical strings and/or conformal field theories usually requires duality conjectures~\cite{carlip2005a,manschot2007a,yin2008a,elshowk2012a,carlip2023a,ma2024a}.

In $ n=4 $ the situation is, of course, worse. Gravity has propagating local degrees of freedom  in the bulk, and any fundamental equivalence to TQFTs is lost. However, TQFTs still present a viable venue for an ultraviolet (UV) completion of gravity. In this scenario, GR is treated as an effective field theory, which emerges out of a TQFT phase after a symmetry break occurs~\cite{sako1997a,mielke2011a,sadovski2017a,gu2017a}. Remarkably, this proposal is able to address several issues of the very early Universe without the need of an inflationary period~\cite{agrawal2020a,kehagias2021a,fang2023a}. And, it can potentially shed light in the nature of dark matter, and hidden supersymmetry~\cite{fang2021a,raitio2024a}.

In this work, we propose an $ n=4 $ $ SO(4) $ TYM theory as a topological phase of gravity. In Section~\ref{sec:top_grav}, we show how it is able to generate not only GR, but the whole family of Lovelock-Cartan theories of gravity\footnote{This is the most general gravitational dynamics in $ n = 4 $, which includes curvature and torsion, but excludes high-derivative terms~\cite{mardones1991a,hassaine2016a,corichi2016a}.}, after its topological symmetry is explicitly broken via the introduction of a mass scale. And, in Section~\ref{sec:quantum}, we show its Ward identities, counterterms, and quantum stability, to all orders in perturbation theory. Finally, in Section~\ref{sec:tym}, we give a brief review on TYM theories, and Section~\ref{sec:conclusions} contain our conclusions.

\section{Topological Yang-Mills theory}\label{sec:tym}

\subsection{Mathematical preliminaries}\label{ssec:math-preliminaries;sec:tym}

Let spacetime be the standard $\mathbb{R}^4$, endowed with the globally flat metric\footnote{The tensor product symbol $\otimes$, its fully symmetrized version $\vee$, and its fully anti-symmetrized version (the wedge product) $\wedge$, will be omitted when the context is sufficiently clear. Greek indexes run from 0 to 3.} $ \mathrm{g} =\delta_{\mu\nu}dx^\mu dx^\nu$. The global moving co-frame $dx^\mu$ diagonalizes $ \mathrm{g} $, $ \partial_{ _\mu } $ is its dual frame, and $d=dx^{\mu}\partial_{ \mu }$ is the exterior derivative on $\mathbb{R}^4$.

Let $G$ be a matrix Lie group, and $\mathfrak{g}$ its Lie algebra with Lie bracket
\begin{equation}
  \label{eq:G lie algebra}
  \left[ J_{A}, J_{B} \right]=\tensor{f}{_{AB}^C}J_C \;.
\end{equation}
$J_A$ is a tangent frame next to the identity in $G$, and $\tensor{f}{_{AB}^C}$ are the structure constants\footnote{Uppercase Latin indexes run from 1 to $\dim\left(G\right)$.}. $G$ is also a smoothable manifold, naturally endowed with a globally flat metric $ \mathrm{k} = \tensor{ f }{ _{AC}^D } \tensor{ f }{ _{BD}^C } J^A J^B $. The co-frame $ J^A $ is dual to $J_A$, and $ \mathrm{k} $ diagonalizes if
\begin{equation}
  \label{eq:orthonormal generators}
  \tr\left(J_A J_B\right) = \frac{1}{2}\delta_{AB} \;.
\end{equation}
This choice also fully anti-symmetrizes $f_{ABC}\equiv \tensor{f}{_{AB}^D}\delta_{DC}$.

Consider another smoothable manifold $ P $, the principal bundle structure $G \hookrightarrow P \rightarrow \mathbb{R}^4$, its adjoint bundles $ \mathrm{Ad} P \equiv P \times_G G $ and $ \mathrm{ad}P \equiv P \times_G \mathfrak{g}$. A gauge field on $ \mathbb{R}^4 $ is an element of $C^\infty \left( \mathrm{ad}P \otimes T^*\mathbb{R}^4 \right)$ --- a $ \mathrm{ad} P $-valued 1-form field, $ A=\tensor{A}{ ^A_\mu }J_A dx^\mu $. It results from the pullback of a $G$-connection living on $ P $. The space of all $G$-connections on $ P $ is $ \mathcal{A} \equiv C^{ \infty } \left( J^1 P \right) $, where $ J^1 P $ is its 1st jet bundle. The space of all gauge transformations is $ \mathcal{G} \equiv C^{ \infty }\left( \mathrm{Ad}P \right) $, and $ C^{ \infty } \left( \mathrm{ad} P \right) $ is its Lie algebra.

The full geometrical arena of a gauge theory is that of the universal bundle $\left( G\times\mathcal{G} \right) \hookrightarrow \left( P \times \mathcal{A} \right) \rightarrow \left( \mathbb{R}^4\times\mathcal{A}/\mathcal{G} \right)$~\cite{baulieu1985a,baulieu1988a}. Analogous to above, a universal gauge field $\tilde{A}$ is the result of a pullback to $\mathbb{R}^4 \times \mathcal{A}/\mathcal{G}$, of a universal $\left( G \times \mathcal{G} \right)$-connection living on $P \times \mathcal{A}$. It can be written as
\begin{equation}
  \tilde{A} = A + c \;,
\end{equation}
where the $\mathrm{ad}P$-valued 0-form $c=c^A J_A$ is a local projection of a Maurer-Cartan form on $ \mathcal{G} $ --- also known as the Faddeev-Popov (FP) ghost field. $A$ and $c$ can be seen as the (1,0) and (0,1) component of $\tilde{A}$, respectively, in relation to the product $\mathbb{R}^4 \times \mathcal{A}/\mathcal{G}$.

The exterior derivative $\tilde{d}$ on $\mathbb{R}^4 \times \mathcal{A}/\mathcal{G}$ can be written as
\begin{equation}\label{eq:universal-exterior-derivative}
  \tilde{d} = d + s \;,
\end{equation}
where $s$ is the exterior derivative on $\mathcal{G}$ --- also known as the BRST operator. The graded exterior algebra defined by $\tilde{d}$ gives the meaning of an 1-form with ghost number 0, and a 0-form with ghost number 1, respectively, to the components $ \left( 1,0 \right) $, and $ \left( 0,1 \right) $. Accordingly, the total grading (\textit{a.k.a.}, statistics) of $A$, and $c$, is odd (\textit{a.k.a.}, fermionic). Additionally, $\tilde{d}$, $d$, and $s$ are all nilpotent operators by definition. Equation~\eqref{eq:universal-exterior-derivative} resumes to $sd+ds=0$, which means $d$ and $s$ are fermionic operators.

The universal curvature $ \tilde{F} $ of $\tilde{A}$ is given by
\begin{subequations}\label{eq:universal-curvature}
  \begin{align}
    \tilde{F} & \equiv \tilde{d}\tilde{A}+\tilde{A}^2\;, \\
              & = F+\psi+\phi \;.
  \end{align}
\end{subequations}
It has the curvature $F\equiv dA+A^2$ of $A$ as the (2,0) component, and the so-called \emph{2nd generation ghost fields}, $\psi \equiv sA + Dc$ and $\phi \equiv sc + c^2$, as the (1,1) and (0,2) components, respectively. Here, $D=d+ \left[ A, \phantom{A} \right]$ is the covariant exterior derivative acting on $\mathrm{ad}P$-valued forms\footnote{From now on, $\left[ \phantom{ A }, \phantom{ A } \right]$ should be seen as a $\tilde{d}$ graded Lie bracket, taking into account the total statistics of each $\mathrm{ad}P$-valued form being input in it.}. The total grading of $F$, $\psi$ and $\phi$ is even (\textit{a.k.a.}, bosonic). The grading of all the fields introduced so far can be found in Table~\ref{tab:tym-grading}. Bianchi identities for $\tilde{F}$, and $F$, give
\begin{subequations}\label{eq:universal-bianchi-identity}
  \begin{align}
    \tilde{D}\tilde{F}          & =0 \;,  \\
    sF+D\psi+\left[ c,F \right] & = 0 \;.
  \end{align}
\end{subequations}

\subsection{Symmetries, observables and dynamics}\label{ssec:sym-and-obs;sec:tym}

The traditional Yang-Mills (YM) BRST symmetry transformations,
\begin{subequations}\label{eq:ym-brst}
  \begin{align}
    s_{\text{YM}}A & = -Dc \;,                   \\
    s_{\text{YM}}c & = - c^2 \;,                 \\
    s_{\text{YM}}F & = - \left[ c, F \right] \;,
  \end{align}
\end{subequations}
can be obtained from~\eqref{eq:universal-curvature}, and~\eqref{eq:universal-bianchi-identity}, by enforcing the so-called \emph{horizontal condition}, $\psi=\phi=0$. However, the full gauge structure described in Section~\ref{ssec:math-preliminaries;sec:tym}, in general, leads to a much stronger set of symmetry transformations. Explicitly,
\begin{subequations}\label{eq:tym-brst}
  \begin{align}
    sA    & = -Dc + \psi \;,                      \\
    sc    & = - c^2 + \phi \;,                    \\
    s\psi & = -D\phi - \left[ c, \psi \right] \;, \\
    s\phi & = - \left[ c, \phi \right]\;,         \\
    sF    & = -D\psi - \left[ c, F \right] \;.
  \end{align}
\end{subequations}
This is known as the Topological Yang-Mills (TYM) BRST transformations~\cite{baulieu1988a}.

The $ s $-cohomology forbids the presence of the traditional YM Lagrangian density, $ \tr \left( F \star F \right) $ --- where $ \star $ is the Hodge dual\footnote{ $ \star F = \frac{ \sqrt{ |\mathrm{g}| } }{ 4 } \tensor{ \epsilon }{ _{\mu \nu}^{\alpha \beta} } F_{\alpha \beta} dx^\mu dx^\nu $. } on $ \mathbb{R}^4 $. In fact, it forbids the presence of any $ \mathrm{g} $~metric-contaminated observable. In the absence of spontaneous symmetry breaking, all non-trivial observables are elements in the $s$-cohomology modulo $d$-boundaries. The only non-trivial ones allowed by~\eqref{eq:tym-brst} are of the type
\begin{equation}\label{eq:tym-observables}
  \mathcal{O}_k = \tr \left( \tilde{F}^k \right) \; ; \; k \in {\mathbb{N}}^{\ge 1} \;.
\end{equation}
These invariant polynomials of $\tilde{F}$ can be used to construct the Chern classes of the universal bundle. In other words, they are all topological in nature. In particular,
\begin{equation}\label{eq:donaldson-polynomials}
  \mathcal{O}_2 = \tr \left[ F^2 + 2\psi F + \left( 2\phi F + \psi^2 \right) + 2\psi \phi + \phi^2\right]
\end{equation}
contains, precisely, the Donaldson polynomials evaluated in the seminal works of S.~K.~Donaldson~\cite{donaldson1983a,donaldson1990a}, and E.~Witten~\cite{witten1988d}.

Among all allowed observables in~\eqref{eq:tym-observables}, only the (4,0) component of~\eqref{eq:donaldson-polynomials} is suitable for a Lagrangian density at four spacetime dimensions. Thus, the TYM action functional is defined to be
\begin{equation}\label{eq:tym-action}
  S_{\text{TYM}}\left[ A \right] \equiv  \int\tr\left( g F^2\right) \;,
\end{equation}
where $ g $ is a dimensionless coupling parameter. The Lagrangian density $\tr \left(g F^2 \right)$ is proportional to the 1st Pontryagin number of $\mathbb{R}^4$. And, its integral is proportional to the (compactly supported) Hirzebruch signature of $\mathbb{R}^4$. In other words,~\eqref{eq:tym-action} is a topological  invariant of spacetime.

The field equations are trivial ($0=0$), which signals a lack of local dynamics. Instead, TYM has a non-trivial non-local dynamics in the bulk. Many topological field theories can be formulated as fully extended functorial field theories~\cite{atiyah1988a,segal1988a,baez1995a,schreiber2009a,baez2009a}. In this context, their non-local bulk dynamics can be roughly understood as the propagation and scattering of topologically embedded fully extended cobordisms.

\subsection{Quantum properties}\label{ssec:quantum-properties;sec:tym}

A partition function for TYM, formally defined in the weakly coupled regime, requires~\eqref{eq:tym-action} to be gauge fixed. Quantum TYM (QTYM) has been studies in several different gauge choices~\cite{baulieu1988a,myers1990c,brandhuber1994a,piguet1995a,sadovski2017c,sadovski2018a,sadovski2018b}. Here, we adopt the (anti-){}self-dual Landau ((A){}SDL) conditions
\begin{subequations}\label{eq:asdlg}
  \begin{align}
    d \star A    & = 0 \;, \\
    d \star \psi & = 0 \;, \\
    F^{\pm}      & = 0 \,,
  \end{align}
\end{subequations}
where $ F^{\pm} \equiv F \pm \star F $.

The $ s $-cohomology guarantees that QTYM is free of gauge anomalies~\cite{baulieu1988a}. The (A){}SDL gauge choice is convenient because it results in a very strong set of Ward identities. In this gauge, QTYM is shown to be renormalizable to all orders in perturbation theory. It has only one independent, and non-physical, renormalization~\cite{sadovski2017c}. Moreover, all connected $n$-point Green functions are tree-level exact~\cite{sadovski2018a}. Clearly, this gauge makes evident that QTYM also has no local degrees of freedom. And, due to the lack of loop corrections, the classical observables in~\eqref{eq:tym-observables} maintain their topological nature: the QTYM observables are still Donaldson polynomials.

Gauge conditions are, generally, $ \mathrm{g} $ metric-contaminated. This is evident in~\eqref{eq:asdlg}, due to the presence of the $\star$ operator. Consequently, care is needed to implement them without spoiling the topological nature of the partition function. The standard procedure, in the BRST quantization scheme, is to introduce Lautrup-Nakanishi fields --- a pair for each gauge condition --- satisfying the $s$-doublet condition,
\begin{subequations}\label{eq:anti-ghost_lautrup-nakanishi}
  \begin{align}
    s \bar{ c }  & = b \;, \;\; s b  = 0 \;,                    \\
    s \bar{\chi} & = B \;, \;\; s B        = 0 \;,              \\
    s \bar{\phi} & = \bar{\eta} \;, \;\; s \bar{ \eta } = 0 \;.
  \end{align}
\end{subequations}
We refer to Table~\ref{tab:tym-grading} for the grading of these fields.

The doublet theorem guarantees that no observable can be made out of these quantities --- terms containing $\bar{c}$, $b$, $\bar{\chi}$, $B$, $\bar{\phi}$, and/or $\bar{\eta}$ are, at most, $s$-boundaries. Following this spirit, we define the (A){}SDL gauge fixing action as
\begin{subequations}\label{eq:tym-gf-action}
  \begin{align}
    S_{\text{GF}} & = s \int \tr \left[ \bar{c} d \star A + \bar{\phi} d \star \psi + \bar{\chi} F^{\pm} \right] \;,                                                                                                                                 \\
                  & =  \int \tr \left[ b d \star A - \bar{c} d \star Dc + \left( \bar{c} + \bar{\eta} + \left[ c,\bar{\phi} \right] \right) d \star \psi + \bar{\phi} d \star D \phi + d c \left[ \star \psi ,\bar{\phi} \right] + \right. \nonumber \\
                  & + \left. \left( B + \left[ c,\bar{\chi} \right] \right) F^{\pm} + \bar{\chi} {\left( D \psi \right)}^{ \pm } \right] \;.
  \end{align}
\end{subequations}

\begin{table}[htpb]
  \caption{Grading of all TYM fields.}%
  \label{tab:tym-grading}
  \begin{tabular}{cccccccccccccc}
    \toprule
    Field      & $A$ & $F$  & $c$ & $\psi$ & $\phi$ & $\bar{c}$ & $b$  & $\bar{\chi}$ & $B$  & $\bar{\phi}$ & $\bar{\eta}$ \\
    \midrule
    Form rank  & 1   & 2    & 0   & 1      & 0      & 0         & 0    & 2            & 2    & 0            & 0            \\
    Ghost no.  & 0   & 0    & 1   & 1      & 2      & -1        & 0    & -1           & 0    & -2           & -1           \\
    Statistics & odd & even & odd & even   & even   & odd       & even & odd          & even & even         & odd          \\
    \bottomrule
  \end{tabular}
\end{table}

\section{Topological phase of gravity}%
\label{sec:top_grav}

\subsection{\texorpdfstring{$SO(4)$}{SO(4)} TYM theory}
\label{ssec:so4tym;sec:top_grav}

Gauge descriptions of gravity are usually associated with a principal frame bundle $G \hookrightarrow Fr \rightarrow \mathbb{R}^4 $. Frequently, the assumption $G = GL^{+}\left( 4, \mathbb{R} \right)$ is considered. Additionally, any smoothable manifold accepts Riemannian structure. By choosing only orthogonal frames with respect to it, one effectively contracts $GL^{+} \left( 4,\mathbb{R} \right)$ down to its orthogonal subgroup $SO \left( 4 \right)$. Unfortunately, not every smoothable 4-manifold accepts a Lorentzian structure. The analogue procedure, resulting in the more physical $SO \left( 1,3 \right)$ gauge theory, is topology-dependent.

Here, we consider only non-compact smoothable 4-manifolds as they are guaranteed to have either Riemannian or Lorentzian structure. The particular choice is irrelevant for TYM theory since its observables are $ \mathrm{g} $~metric-independent: Riemannian and Lorentzian TYM define the same physical theory. For convenience, we adopt the positive-definite $ \mathrm{g} $.

Let $\sigma_{ab}=-\sigma_{ba}$ be the 6 linearly independent generators\footnote{Lowercase Latin indexes run from 0 to 3.} of the Lie algebra $\mathfrak{so}\left( 4 \right)$ of $SO(4)$. They are chosen such that
\begin{subequations}%
  \label{eq:so4algebra}
  \begin{align}
    \left[ \sigma_{ab}, \sigma_{cd} \right] & =-8\tensor{\delta}{^{e}_{\left[ c \right.}}\tensor{\delta}{_{\left. d \right] \left[ a \right.}} \tensor{\delta}{_{\left. b \right]}^f} \sigma_{ef} \;, \\ 
    \tr (\sigma_{ab}\sigma_{cd})            & = 4 \tensor{\delta}{_{ a \left[ c \right. }} \tensor{\delta}{_{ \left. d \right] b }} \;.                                                               
  \end{align}
\end{subequations}
It is commonplace in the literature to denote by $\omega=\tensor{\omega}{^{ab}_\mu} \sigma_{ab}dx^\mu$ the $\mathrm{ad} Fr $-valued connection 1-form, and by $R=d \omega + \omega^2$ its curvature 2-form. We stress that $ \omega $ and $ R $ are, respectively, the same mathematical objects as $A$ and $F$, defined in Section~\ref{sec:tym}.

There are two particular consequences of having $G=SO(4)$, which is relevant to us. And, these are direct results of the Levi-Civita permutation symbol, $ \epsilon_{a_1 \ldots a_N} $, being an $ SO(N) $ invariant tensor --- a statement which is not true for any $G$. First, the Pfaffian of $R$ is, now, a well-defined and non-vanishing quantity,
\begin{subequations}%
  \label{eq:euler-class}
  \begin{align}
    \pf \left( R \right) & \equiv \frac{ 1 }{ 8 } \epsilon_{abcd} R^{ab}R^{cd} \;, \\
                         & =\frac{ 1 }{ 16 } \tr \left( RR^* \right) \;,
  \end{align}
\end{subequations}
where $*$ is the Hodge dual\footnote{ $ R^* = \frac{ 1 }{ 2 } R^{ab}\tensor{ \epsilon }{_{ab}^{cd}} \sigma_{cd} $. } on $SO(4)$. This is equivalent to define the Euler class of $ Fr $. Second, the existence of $*$ also allows us to extend the notion of (anti-){}self duality to
\begin{equation}
  \label{eq:extended-self-duality}
  R^{ \pm } = 0  \;,
\end{equation}
where $ R^{ \pm } \equiv R \pm \star \left( R + R^{ * } + \star R^{ * } \right) $. And, consequently, extend the (A){}SDL gauge conditions to
\begin{subequations}%
  \label{eq:extended-asdlg}
  \begin{align}
    d \star \omega & = 0 \;, \\
    d \star \psi   & = 0 \;, \\
    R^{ \pm }      & =0 \;.
  \end{align}
\end{subequations}
We choose to work in this extended gauge as it improves the renormalizability behavior of the model we are about to construct --- see Section~\ref{sec:quantum}.

The BRST transformations remain the same, of course,
\begin{subequations}%
  \label{eq:top_grav_brst}
  \begin{align}
    s\omega & = -Dc + \psi \;,                      \\
    sc      & = - c^2 + \phi \;,                    \\
    s\psi   & = -D\phi - \left[ c, \psi \right] \;, \\
    s\phi   & = - \left[ c, \phi \right]\;,         \\
    sR      & = -D\psi - \left[ c, R \right] \;.
  \end{align}
\end{subequations}
In particular, the observables are the same, except for the presence of~\eqref{eq:euler-class}.

We define our topological symmetry-restored phase of gravity via the action functional
\begin{equation}
  \label{eq:top_grav_action}
  S_{\text{TG}}\left[ \omega \right] \equiv \int \tr \left( g_1 R^2 + g_2 RR^* \right)\;,
\end{equation}
where $g_1$ and $g_2$ are dimensionless coupling parameters. Of course, couplings involving $ \star $ are forbidden by the topological BRST symmetry~\eqref{eq:top_grav_brst}. Its connection to gravity will become clear in Section~\ref{ssec:from_top_to_grav;sec:top_grav}. For now, we restrict ourselves to comment that $S_{\text{TG}}$ is the most general action functional that is:
\begin{enumerate}[label=\roman*)] 
  \item an invariant polynomial of $\omega$ and its derivatives;
  \item local\footnote{The integrand is a function of a single spacetime point. };
  \item power-counting renormalizable;
  \item fully topological --- it is the sum of the (compactly supported) Hirzebrunch signature and the (compactly supported) Euler characteristic of spacetime;
  \item  and, by definition, an $s$-cycle which is not an $s$-boundary.
\end{enumerate}

\subsection{Adding a BRST boundary}%
\label{ssec:brst_boundary;sec:top_grav}

If we add an $s$-boundary to~\eqref{eq:top_grav_action}, we strictly define a new dynamics. However, since $s$-boundaries lie outside the $s$-cohomology groups, the set of observables remains unchanged. In other words, we are still describing the same physical system. This is exactly what we did in Section~\ref{ssec:quantum-properties;sec:tym} to gauge fix TYM theory.\@ However, our objective here is not to gauge fix any symmetries, but to employ Symanzik's technique~\cite{symanzik1970a}. Consider the pair of classical external fields, $X$ and $Y$, satisfying the $s$-doublet condition
\begin{equation}
  \label{eq:s-doublet}
  sY = X \;, \;\; sX = 0 \;.
\end{equation}
Their grading are displayed in Table~\ref{tab:grading2}. And, the most general action functional that is:
\begin{enumerate}[label=\roman*)] 
  \item an invariant polynomial of $\omega$, $X$, $Y$ and their derivatives;
  \item local;
  \item power-counting renormalizable;
  \item  and, an $s$-boundary;
\end{enumerate}
is given by
\begin{subequations}%
  \label{eq:s-boundary_action}
  \begin{align}
    S_{\text{sb}} & = s \int \tr \left[ Y \left( g_3 R + g_4 \star R + g_5 R^* + g_6 \star R^* + g_7 X + g_8 \star X + g_9 X^* \right. + \right.                                                                \nonumber                                               \\
                  & \left. + \left. g_{10} \star X^* \right) \right] \;,                                                                                                                                                                                                \\
                  & = \int \tr \left\{ \left( g_3 R + g_4 R \star + g_5 R^* + g_6 R^* \star + g_7 X + g_8 X \star + g_9 X^* + g_{10} X^* \star \right) X \right. +                                                       \nonumber                                      \\
                  & + Y \left[ g_3 \left( D \psi + \left[ c , R \right] \right) + g_4 \star \left( D \psi + \left[ c , R \right] \right) + g_5 {\left( D \psi + \left[ c , R \right] \right)}^* + g_6 \star \left( D \psi \right. + \right.                   \nonumber \\
                  & \left. + \left. { \left. \left[ c , R \right] \right) }^* \right] \right\} \;,                                                                                                                                                                      
  \end{align}
\end{subequations}
where all the $g_i$ coupling parameters are dimensionless.

The action $S_{\text{sb}}$ is not a topological invariant. Nevertheless, we emphasize that the theory defined by
\begin{equation}
  \label{eq:tg+sb}
  S=S_{\text{TG}}+S_{\text{sb}}
\end{equation}
still is a topological field theory. And, it is physically indistinguishable from $S_{\text{TG}}$.
\begin{table}[htpb]
  \caption{Grading of Symanzik sources.}%
  \label{tab:grading2}
  \begin{tabular}{ccc}
    \toprule
    Field      & $X$  & $Y$ \\
    \midrule
    Form rank  & 2    & 2   \\
    Ghost no.  & 0    & -1  \\
    Statistics & even & odd \\
    \bottomrule
  \end{tabular}
\end{table}

\subsection{From topology to gravity}%
\label{ssec:from_top_to_grav;sec:top_grav}

Before finally proceeding to the connection between our topological model and gravitational field theories, it is convenient to acknowledge that, in the presence of a $ \mathrm{g} $ structure, a Clifford bundle $ \mathrm{Cl} Fr= Fr \times_{SO(4)}Cl_4 \left( \mathbb{R} \right)$ can be associated to $ Fr $. A typical moving frame, $\left\{ \mathds{1}_4, \gamma_a, \tensor{ \gamma }{ _{ \left[ a \right. } } \tensor{ \gamma }{ _{ \left. b \right] } }, \gamma_5, \gamma_5 \gamma_a \right\}$, consists of 16 matrices such that $ \left\{ \gamma_a, \gamma_b \right\} = 2 \delta_{ab}$, and $\gamma_5 \equiv \gamma_0 \gamma_1 \gamma_2 \gamma_3 $. In particular, $\tensor{ \gamma }{ _{ \left[ a \right. } } \tensor{ \gamma }{ _{ \left. b \right] } }$ is an $\mathfrak{so}\left( 4 \right)$ representation satisfying~\eqref{eq:so4algebra}. The convenience here is that differential forms valued in the adjoint and fundamental representation space of $SO(4)$ are treated in the same footing --- they are both $ \mathrm{Cl}Fr $-valued differential forms, also known as Clifforms~\cite{benn1987a,mielke2001a,mielke2017a}. 

Gravity is described by a special kind of gauge theory in the sense that $ \mathrm{Cl}Fr $ is isomorphic to the Clifford bundle of spacetime. The isomorphism is given by $\gamma_a = \tensor{ e }{ _a^\mu } \gamma_\mu $, where $e^a \equiv \tensor{ e }{ ^a_\mu } dx^\mu $ is the $\mathrm{Cl}Fr$-valued soldering 1-form --- also known as the vierbein field ---, and $ e_a \equiv \tensor{ e }{ _a^\mu }\partial_{ \mu }$ is its dual vector. Finally, for latter use, we define $ \gamma \equiv \gamma_a e^a $. And, it is straightforward to show that $\star \gamma^2 = {\left( \gamma^2 \right)}^*= \gamma_5 \gamma^2$.

Returning to the $SO(4)$ TYM theory, the term \textquote{topological symmetry-restored phase of gravity}, used in Section~\ref{ssec:so4tym;sec:top_grav}, was deliberately chosen to suggest that a traditional gravity theory, with propagating local degrees of freedom, can be obtained from~\eqref{eq:tg+sb} via a symmetry breaking mechanism. In particular, an explicit symmetry breaking (ESB) that achieves that is obtained by forcing Symanzik sources, $X$ and $Y$, to attain physical values:
\begin{subequations}%
  \label{eq:physicalxy}
  \begin{align}
    Y\big|_{\text{phys.}} & = 0 \;, \label{eq:physicalsource1}              \\
    X\big|_{\text{phys.}} & = \mu^2 \gamma^2 \;, \label{eq:physicalsource2}
  \end{align}
\end{subequations}
where $\mu$ is a mass scale. Indeed, when $X$ and $Y$ are identified accordingly in~\eqref{eq:tg+sb}, the result
\begin{align}
  \label{eq:llc_action}
  S\big|_{\text{phys.}} & = \int \tr \left\{ g_1 R^2 + g_2 R R^* + \mu^2 \left[ \left( g_4 + g_5 \right) R \star \gamma^2 + \mu^2 \left( g_8 + g_9 \right) \gamma^2 \star \gamma^2  \right. + \right. \nonumber \\
                        & \left. + \left. \left( g_3 + g_6 \right) R \gamma^2 \right] \right\} \;,
\end{align}
can be immediately recognized as the Lovelock-Cartan theory of gravity on non-compact 4-manifolds~\cite{mardones1991a,hassaine2016a,corichi2016a}. The coupling $\tr \left( R \star \gamma^2 \right)$ is the Einstein-Palatini Lagrangian density, $\tr \left( \gamma^2 \star \gamma^2 \right)$ is the cosmological constant term, and $\tr \left( R \gamma^2 \right)$ is the Holst term~\cite{holst1996a} --- related to the torsional Nieh-Yan invariant polynomial~\cite{nieh1982a,chandia1997a,nieh2007a}.

It is important to clarify that the ESB employed here is inspired by the well-established Symanzik source technique~\cite{symanzik1970a}. There are two equivalent ways to approach it. One can start with a symmetry broken theory, \textit{e.g.}, defined by the action~\eqref{eq:llc_action}, then external sources are introduced in order to control it. This effectively embeds the theory into a larger --- more symmetric --- one. After the calculations are performed, the sources can be set to values which explicitly undo the embedding. Just as well, one can start with the larger theory. Then, Symanzik sources are introduced, and their attained value represent an ESB\footnote{We remark that the Symanzik method is employed in a wide range of field theoretical models in which one wishes to control broken symmetries. The Gribov-Zwanziger model~\cite{zwanziger1981a,zwanziger1989a,dudal2005a,vandersickel2012a}, and Lorentz violating models~\cite{sobreiro2015a,sobreiro2016b,sobreiro2017b}, are just a few examples.}.

The field equations for the gravitational fields $\gamma$ and $\omega$ are, respectively,
\begin{subequations}%
  \label{eq:grav_field_eqs}
  \begin{align}
    \left[ \left( g_3 + g_6 \right) R + \left( g_4 + g_5 \right) \gamma_5 R + 2 \mu^2 \left( g_8 + g_9 \right) \gamma_5\gamma^2 , \gamma \right] & = 0 \;, \\
    \left[ \left( g_3 + g_6 \right) T + \left( g_4 + g_5 \right) \gamma_5 T, \gamma \right]                                                      & = 0 \;,
  \end{align}
\end{subequations}
where $ T = D \gamma $ is the $\mathrm{Cl}Fr$-valued torsion 2-form. Their solutions are Riemann-Cartan spacetimes, in general.

The presence of curvature and torsion is due to our adoption of the most general construction in~\eqref{eq:s-boundary_action}. If one wishes to generate Einstein gravity exclusively, the values of the $ g_i $ can be tweaked by hand to do so. We chose not to do so, as it goes against the quantum field theory paradigm. We just determine the symmetries. It is the symmetries that determine all allowed couplings, and their relative relevance through $ \beta $-functions.

Via the Correspondence Principle,
\begin{subequations}%
  \label{eq:correspondence}
  \begin{align}
    \mu^2 \left( g_4+g_5 \right) & = \frac{{m_P}^2}{32\pi} \;,            \\
    \mu^4 \left( g_8+g_9 \right) & = -\frac{{m_P}^2\Lambda^2}{384\pi} \;,
  \end{align}
\end{subequations}
where ${m_P}^2$ is Planck mass and $\Lambda^2$ the cosmological constant. The LHS of~\eqref{eq:correspondence} contain coupling parameters coming from a well-defined perturbative quantum field theory. In principle, their renormalized values can be obtained, thus predicting the values of the Planck scale, and the size of the observable universe. This will be the topic of a future work.

In Section~\ref{ssec:so4tym;sec:top_grav}, we commented on how the traditional YM BRST symmetry, given by~\eqref{eq:ym-brst}, is obtained from the TYM BRST, given by~\eqref{eq:tym-brst}, via the horizontal condition: $ s_{\text{YM}} = s|_{\psi = \phi = 0} $. Let $ s = s_{\text{YM}} + s_{\text{T}} $, where $ s_{\text{T}} $ is the \textquote{topological sector of $s$} given by
\begin{subequations}%
  \label{eq:s_T}
  \begin{align}
    s_{\text{T}} \omega & = \psi \;,                              \\
    s_{\text{T}} c      & = \phi \;,                              \\
    s_{\text{T}} \psi   & = - D \phi - \left[ c, \psi \right] \;, \\
    s_{\text{T}} \phi   & = - \left[ c, \phi \right] \;,          \\
    s_{\text{T}} R      & = - D \psi \;.
  \end{align}
\end{subequations}
The topological symmetry-restored phase of gravity, defined via the action functional \eqref{eq:top_grav_action}, is an $s$-cycle. Meanwhile, the induced (Lovelock-Cartan) gravity, defined via~\eqref{eq:llc_action}, is an $ s_{\text{YM}} $-cycle. Clearly, the ESB above implements the horizontal condition at a dynamical level. It deforms $ s $ into $ s_{\text{YM}} $ by breaking $ s_{\text{T}} $.

The $ s_{\text{YM}} $-cohomology differs from the $ s $-cohomology in a very important way: it allows for local observables. For instance, the YM Lagrangian density $ \tr \left( R \star R \right) $ which, in a gravitational context, is recognizable as the Kretschmann scalar. Ultimately, the ESB~\eqref{eq:physicalxy} can be physically interpreted as responsible for freeing the local degrees of freedom of gravity in the bulk --- originally frozen due to full invariance under $s$.

\section{Algebraic renormalizability}%
\label{sec:quantum}

To prove the renormalizability of the proposed topological symmetry-restored phase of gravity, we follow the algebraic renormalization program~\cite{piguet1995b}. The results obtained are valid to all orders in perturbation theory, and are independent of any regularization scheme.

First, we fix the gauge symmetry of~\eqref{eq:tg+sb} by adding to it the extended (A){}SDL gauge fixing action,
\begin{subequations}%
  \label{eq:tg-gf-action}
  \begin{align}
    S_{\text{GF}} & = s \int \tr \left[ \bar{c} d \star \omega + \bar{\phi} d \star \psi + \bar{\chi} R^{\pm} \right] \;,                                                                                                                                \\
                  & = \int \tr \left[ b d \star \omega - \bar{c} d \star Dc + \left( \bar{c} + \bar{\eta} + \left[ c,\bar{\phi} \right] \right) d \star \psi + \bar{\phi} d \star D \phi + d c \left[ \star \psi ,\bar{\phi} \right] + \right. \nonumber \\
                  & + \left. \left( B + \left[ c,\bar{\chi} \right] \right) R^{\pm} + \bar{\chi} {\left( D \psi \right)}^{\pm} \right] \;.
  \end{align}
\end{subequations}

Second, the non-linearity of the BRST transformations, and of the non-linear bosonic ghost symmetry,~\eqref{eq:nl-bosonic-eq}, will inevitably appear as insertions in the correlation function of the quantized theory. To account for these infinities, we need to explicitly include them in the total action. Consider the $ s $-doublets,
\begin{subequations}%
  \label{eq:nl-s-doublets}
  \begin{align}
    s\tau    & = \Omega \;, \;\; s\Omega   = 0 \;, \\
    sE       & = L \;, \;\; sL        = 0 \;,      \\
    s\lambda & = K \;, \;\; sK        = 0 \;,      \\
    sZ       & = H \;, \;\; sH        = 0 \;.
  \end{align}
\end{subequations}
The non-linearity action functional
\begin{subequations}%
  \label{eq:nl-action}
  \begin{align}
    S_{\text{NL}} & = s \int \tr \left( \tau Dc + E c^2 + \lambda \left[ c,\bar\chi \right] + Z \left[ c,Y \right] \right) \;, \label{eq:nl-action-1}                                                                                           \\
                  & = \int \tr \left\{ \Omega Dc + Lc^2 + \tau \left( D\phi + \left[ c,\psi \right] \right) + E \left[ c,\phi \right] + K \left[ c,\bar\chi \right] + \lambda \left(\vphantom{^2}\left[ c,B \right] + \right. \right. \nonumber \\
                  & + \left. \left. \left[ c^2,\bar\chi \right] + \left[ \bar\chi,\phi \right] \right) + H \left[ c, Y \right] + Z \left( \left[ Y, \phi \right] + \left[ c^2, Y \right] + \left[ c, X \right] \right) \right\} \;,
  \end{align}
\end{subequations}
is the most general local, power-counting renormalizable, invariant polynomial that explicitly couples all independent non-linearities to external sources, while remaining an $ s $-boundary. The grading of the newly introduced fields can be found at Table~\ref{tab:nl-sources}.

The full classical action to be considered is
\begin{equation}
  \label{eq:total-action}
  \Sigma \equiv S + S_{\text{GF}} + S_{ \text{NL} }\; .
\end{equation}

\begin{table}[htpb]
  \caption{Grading of Zwanziger sources for symmetry non-linearities.}%
  \label{tab:nl-sources}
  \begin{tabular}{ccccccccc}
    \toprule
    Field      & $\tau$ & $\Omega$ & $E$ & $L$  & $\lambda$ & $K$  & $Z$ & $H$  \\
    \midrule
    Form rank  & 3      & 3        & 4   & 4    & 2         & 2    & 2   & 2    \\
    Ghost no.  & -2     & -1       & -3  & -2   & -1        & 0    & -1  & 0    \\
    Statistics & odd    & even     & odd & even & odd       & even & odd & even \\
    \bottomrule
  \end{tabular}
\end{table}

\subsection{Ward identities}%
\label{ssec:ward_identities}
The local symmetries of $ \Sigma $, including linearly broken ones, are:
\begin{itemize}
  \item Traditional gauge fixing equation
        \begin{equation}
          \label{eq:gfeq}
          \frac{ \delta \Sigma }{ \delta b } = d \star \omega  \;;
        \end{equation}
  \item The topological gauge fixing equation
        \begin{equation}
          \label{eq:topgfeq}
          \frac{\delta\Sigma}{\delta\bar{\eta}} = d\star{\psi} \;;
        \end{equation}
  \item The Faddeev-Popov anti-ghost equation
        \begin{equation}
          \label{eq:fpantighosteq}
          \mathcal{G}_{\bar{c}} \left(\Sigma\right)=d\star{\psi} \;,
        \end{equation}
        where
        \begin{equation}
          \label{eq:fpantighostop}
          \mathcal{G}_{\bar{c}}\equiv \frac{\delta\phantom{c}}{\delta\bar{c}}+d\star{\frac{\delta\phantom{\Omega}}{\delta\Omega}} \;;
        \end{equation}
  \item The bosonic anti-ghost equation
        \begin{equation}
          \label{eq:bosonicantighosteq}
          \mathcal{G}_{\bar{\phi}} \left(\Sigma\right)=0\;,
        \end{equation}
        where
        \begin{equation}
          \label{eq:bosonicantighostop}
          \mathcal{G}_{\bar{\phi}}\equiv \frac{\delta}{\delta\bar{\phi}}-d\star{\frac{\delta}{\delta\tau}}\;.
        \end{equation}
\end{itemize}
The equations~\eqref{eq:topgfeq} and~\eqref{eq:fpantighosteq} can be combined into an exact symmetry of $ \Sigma $. Additionally, the global symmetries of $ \Sigma $, including linearly broken ones, are:
\begin{itemize}
  \item  The Slavnov-Taylor identity
        \begin{equation}
          \label{eq:st-identity}
          \mathcal{S}\left(\Sigma\right)=0 \;,
        \end{equation}
        where
        \begin{align}
          \label{eq:SToperator}
          \mathcal{S} & \equiv \int \tr\left[\left(\psi-\frac{\delta\phantom{\Omega}}{\delta\Omega}\right)\frac{\delta\phantom{\omega }}{\delta \omega }+\left(\phi-\frac{\delta\phantom{L}}{\delta L}\right)\frac{\delta\phantom{c}}{\delta c}-\frac{\delta\phantom{\tau}}{\delta\tau}\frac{\delta\phantom{\psi}}{\delta\psi}-\frac{\delta\phantom{E}}{\delta E}\frac{\delta\phantom{\phi}}{\delta\phi}+X\frac{\delta\phantom{Y}}{\delta Y} \right. + \nonumber \\
                      & + \left. b\frac{\delta\phantom{\bar{c}}}{\delta\bar{c}}+B\frac{\delta\phantom{\bar{\chi}}}{\delta\bar{\chi}}+\bar{\eta}\frac{\delta\phantom{\bar{\phi}}}{\delta\bar{\phi}} + \Omega \frac{\delta\phantom{\tau}}{\delta\tau}+L\frac{\delta\phantom{E}}{\delta E}+K\frac{\delta\phantom{\lambda}}{\delta\lambda}+H\frac{\delta\phantom{Z}}{\delta Z}\right] \;;
        \end{align}
  \item The 1st FP ghost equation
        \begin{equation}
          \label{eq:1stfpghosteq}
          \mathcal{G}^{\left(1\right)}_c \left(\Sigma\right)=\Delta_c \;,
        \end{equation}
        where
        \begin{align}
          \label{eq:1stfpghostop}
          \mathcal{G}^{\left(1\right)}_c & \equiv \int \left(\frac{\delta\phantom{c}}{\delta c}-\left[Y,\frac{\delta\phantom{X}}{\delta X}\right]-\left[\bar{c},\frac{\delta\phantom{b}}{\delta b}\right]-\left[\bar{\phi},\frac{\delta\phantom{\bar{\eta}}}{\delta\bar{\eta}}\right]-\left[\bar{\chi},\frac{\delta\phantom{B}}{\delta B}\right]-\left[\lambda,\frac{\delta\phantom{K}}{\delta K}\right] + \right. \nonumber \\
                                         & \left. - \left[Z,\frac{\delta\phantom{H}}{\delta H}\right]\right) \;,
        \end{align}
        and
        \begin{align}
          \label{eq:linearbreakc}
          \Delta_c & \equiv \int \left(\left[\omega ,\Omega\right]+\left[L,c\right]+\left[\tau,\psi\right]+\left[E,\phi\right]+\left[\bar{\chi},K\right]+\left[\lambda,B\right] + \left[Y,H\right] \right. \nonumber \\
                   & + \left. \left[Z,X\right] \right) \;;
        \end{align}
  \item The 2nd Faddeev-Popov ghost equation
        \begin{equation}
          \label{eq:2ndfpghosteq}
          \mathcal{G}^{\left(2\right)}_c \left(\Sigma\right)=\Delta_c \;,
        \end{equation}
        where
        \begin{equation}
          \label{eq:2ndfpghostop}
          \mathcal{G}^{\left(2\right)}_c \equiv \int \left(\frac{\delta\phantom{c}}{\delta c}+\left[\omega ,\frac{\delta\phantom{\psi}}{\delta \psi}\right]+\left[c,\frac{\delta\phantom{\phi}}{\delta \phi}\right]-\left[\bar{\phi},\frac{\delta\phantom{\bar{c}}}{\delta\bar{c}}\right]+\left[\tau,\frac{\delta\phantom{\Omega}}{\delta \Omega}\right]+\left[E,\frac{\delta\phantom{L}}{\delta L}\right]\right)\;;
        \end{equation}
  \item The bosonic ghost equation
        \begin{equation}
          \mathcal{G}_\phi \left(\Sigma\right) = \Delta_\phi \;,
        \end{equation}
        where
        \begin{equation}
          \mathcal{G}_\phi \equiv \int \left(\frac{\delta\phantom{\phi}}{\delta\phi}-\left[\bar{\phi},\frac{\delta\phantom{b}}{\delta b}\right]\right) \;,
        \end{equation}
        and
        \begin{equation}
          \label{eq:linearbreakphi}
          \Delta_\phi \equiv \int \left(\left[\omega ,\tau\right]+\left[c,E\right]+\left[\bar{\chi},\lambda\right]+\left[Y,Z\right]\right) \;;
        \end{equation}
  \item  The vectorial supersymmetry
        \begin{equation}
          \label{eq:vecsusyeq}
          \mathcal{W}\left(\Sigma\right)=0 \;,
        \end{equation}
        where
        \begin{align}
          \label{eq:vecsusyop}
          \mathcal{W} & \equiv \int \tr \left[\mathcal{L}_\xi \omega \frac{\delta\phantom{\psi}}{\delta\psi}+\mathcal{L}_\xi c\frac{\delta\phantom{\phi}}{\delta\phi} - \mathcal{L}_\xi Y \frac{\delta\phantom{X}}{\delta X}-\mathcal{L}_\xi\left(\bar{c}+\bar{\eta}\right)\frac{\delta\phantom{b}}{\delta b} - \mathcal{L}_\xi \bar{\chi}\frac{\delta\phantom{B}}{\delta B}+ \right. \nonumber \\
                      & + \left. \mathcal{L}_\xi\bar{\phi}\left(\frac{\delta\phantom{\bar{c}}}{\delta\bar{c}}-\frac{\delta\phantom{\bar{\eta}}}{\delta\bar{\eta}}\right) + \mathcal{L}_\xi\tau\frac{\delta\phantom{\Omega}}{\delta\Omega}+\mathcal{L}_\xi E\frac{\delta\phantom{L}}{\delta L}+\mathcal{L}_\xi \lambda\frac{\delta\phantom{K}}{\delta K}\right] \;,
        \end{align}
        and $ \mathcal{L}_{\xi} $ is the Lie derivative along the vector field $ \xi $. It is assumed that $\xi$ is a Killing vector, and that it generates a flow of diffeomorphism on spacetime\footnote{Consequently, $\int \mathcal{L}_{\xi} \left( \varphi \star \mathds{1} \right) = 0 \;, \forall \; \varphi \in C^{ \infty }\left( \mathbb{R}^4 \right)$, and $ \star \mathcal{L}_\xi = \mathcal{L}_\xi \star $.}.
  \item  The non-linear bosonic symmetry
        \begin{equation}
          \label{eq:nl-bosonic-eq}
          \mathcal{T}\left(\Sigma\right)=0\;,
          \;
        \end{equation}
        where
        \begin{equation}
          \label{eq:t-operator}
          \mathcal{T} \equiv \int \tr \left[\frac{\delta\phantom{\Omega}}{\delta \Omega}\frac{\delta\phantom{\psi}}{\delta \psi} + \frac{\delta\phantom{H}}{\delta H}\frac{\delta\phantom{X}}{\delta X}+\frac{\delta\phantom{L}}{\delta L}\frac{\delta\phantom{\phi}}{\delta \phi}+\frac{\delta\phantom{K}}{\delta K}\frac{\delta\phantom{B}}{\delta B}+\left(\bar{c}+\bar{\eta}\right)\left(\frac{\delta\phantom{\bar{c}}}{\delta\bar{c}}-\frac{\delta\phantom{\bar{\eta}}}{\delta\bar{\eta}}\right)\right] \;;
        \end{equation}
  \item And, finally, the fermionic ghost symmetry
        \begin{equation}
          \label{eq:fermionicghostsymmetry}
          \mathcal{F}\left(\Sigma\right)=0 \;,
        \end{equation}
        where
        \begin{equation}
          \label{eq:fermionicghostop}
          \mathcal{F} \equiv \int \tr \left[c\frac{\delta\phantom{\phi}}{\delta \phi}+\bar{\phi}\left(\frac{\delta\phantom{\bar{c}}}{\delta\bar{c}}-\frac{\delta\phantom{\bar{\eta}}}{\delta\bar{\eta}}\right)-\tau\frac{\delta\phantom{\Omega}}{\delta \Omega}-2E\frac{\delta\phantom{L}}{\delta L}-\lambda\frac{\delta\phantom{K}}{\delta K}\right] \;.
        \end{equation}
\end{itemize}
Equations~\eqref{eq:1stfpghosteq} and~\eqref{eq:2ndfpghosteq} can also be joined to form an exact symmetry of $ \Sigma  $. The vectorial supersymmetry~\eqref{eq:vecsusyeq} is present in several topological field theories. It is a very strong symmetry, responsible for the 1-loop exactness of $ n=3 $ Chern-Simons theory~\cite{blasi1991a,maggiore1992a,piguet1995a}, and the tree-level exactness of TYM theory~\cite{brandhuber1994a,sadovski2017c,sadovski2018a}, for instance. The non-linear bosonic~\eqref{eq:nl-bosonic-eq}, and the fermionic ghost symmetry~\eqref{eq:fermionicghostsymmetry}, first reported by the authors (and collaborators) in QTYM~\cite{sadovski2017c}, are also present here, and are known to drastically reduce the number of independent renormalizations.

\subsection{Counterterms}%
\label{ssec:counterterm}

The Quantum Action Principle establishes the formal relationship between the set of Ward identities for $ \Sigma $, and the set of Ward identities for its associated quantum effective action, $ \Gamma $. The list above translates to the following set of symmetries for the quantum symmetry-restored phase of gravity:
\begin{subequations}%
  \label{eq:ward-identities}
  \begin{align}
    \frac{ \delta \Sigma^{ \left( n \right) } }{ \delta b }            & = 0 \;,                                \\
    \frac{ \delta \Sigma^{ \left( n \right) } }{ \delta \bar{ \eta } } & = 0 \;,                                \\
    \mathcal{G}_{ \bar{ c } } \left( \Sigma ^{ (n) } \right)           & = 0 \;,                                \\
    \mathcal{G}_{ \bar{ \phi  } } \left( \Sigma ^{ (n) } \right)       & = 0 \;,                                \\
    \frac{ \delta \Sigma^{ \left( n \right) } }{ \delta \bar{ \eta } } & = 0 \;,                                \\
    \mathcal{S}_{\Gamma^{(n-1)}} \left( \Sigma^{ (n)} \right)          & = 0 \;, \label{eq:quantum-st-identity} \\
    \mathcal{G}_{ \phi } \left( \Sigma ^{ (n) } \right)                & = 0 \;,                                \\
    \mathcal{G}^{(1)}_{ c } \left( \Sigma ^{ (n) } \right)             & = 0 \;,                                \\
    \mathcal{G}^{(2)}_{ c } \left( \Sigma ^{ (n) } \right)             & = 0 \;,                                \\
    \mathcal{W} \left( \Sigma^{ (n)} \right)                           & = 0 \;,                                \\
    \mathcal{T}_{\Gamma^{(n-1)}} \left( \Sigma^{ (n)} \right)          & = 0 \;,                                \\
    \mathcal{F} \left( \Sigma^{ (n)} \right)                           & = 0 \;,
  \end{align}
\end{subequations}
where $\Gamma^{(n)} \equiv \Sigma + \epsilon \Sigma^{(1)} + \cdots + \Sigma^{ (n) }$ is $ \Gamma  $ truncated at $ n $-th loop order, $ \epsilon $ is a small perturbative parameter, and $ \Sigma^{ (n)} $ is the $n$-loop radiative correction to $ \Sigma  $. The linear breaks are gone, and the non-linear operators $ \mathcal{S} $ and $ \mathcal{T} $ are replaced by their linearized versions
\begin{align}
  \label{eq:linear-st-operator}
  \mathcal{S}_{\Gamma^{ (n-1) }} & \equiv \int \tr \left[\left(\psi-\frac{\delta\Gamma^{ (n-1) }}{\delta\Omega}\right)\frac{\delta\phantom{\omega}}{\delta \omega}-\frac{\delta\Gamma^{ (n-1) }}{\delta \omega}\frac{\delta\phantom{\Omega}}{\delta \Omega}+\left(\phi-\frac{\delta\Gamma^{ (n-1) }}{\delta L}\right)\frac{\delta\phantom{c}}{\delta c} \right. + \nonumber                                                                                                 \\
                                 & - \left. \frac{\delta\Gamma^{ (n-1) }}{\delta c}\frac{\delta\phantom{L}}{\delta L}-\frac{\delta\Gamma^{ (n-1) }}{\delta\tau}\frac{\delta\phantom{\psi}}{\delta\psi}+\frac{\delta\Gamma^{ (n-1) }}{\delta \psi}\frac{\delta\phantom{\tau}}{\delta \tau}-\frac{\delta\Gamma^{ (n-1) }}{\delta E}\frac{\delta\phantom{\phi}}{\delta\phi} - \frac{\delta\Gamma^{ (n-1) }}{\delta \phi}\frac{\delta\phantom{E}}{\delta E} \right. + \nonumber \\
                                 & + \left. X\frac{\delta\phantom{Y}}{\delta Y}+b\frac{\delta\phantom{\bar{c}}}{\delta\bar{c}}+B\frac{\delta\phantom{\bar{\chi}}}{\delta\bar{\chi}}+\bar{\eta}\frac{\delta\phantom{\bar{\phi}}}{\delta\bar{\phi}}+\Omega\frac{\delta\phantom{\tau}}{\delta\tau}+L\frac{\delta\phantom{E}}{\delta E}+K\frac{\delta\phantom{\lambda}}{\delta\lambda}+H\frac{\delta\phantom{Z}}{\delta Z}\right]\;,
\end{align}
and
\begin{align}
  \label{eq:linear-t-operator}
  \mathcal{T}_{\Gamma^{ (n-1) }} & \equiv \int \tr \left[\frac{\delta\Gamma^{ (n-1) }}{\delta \Omega}\frac{\delta\phantom{\psi}}{\delta \psi} + \frac{\delta\Gamma^{ (n-1) }}{\delta\psi}\frac{\delta\phantom{\Omega}}{\delta \Omega} + \frac{\delta\Gamma^{ (n-1) }}{\delta H}\frac{\delta\phantom{X}}{\delta X} + \frac{\delta\Gamma^{ (n-1) }}{\delta X}\frac{\delta\phantom{H}}{\delta H} \right. + \nonumber \\
                                 & + \left.\frac{\delta\Gamma^{ (n-1) }}{\delta L}\frac{\delta\phantom{\phi}}{\delta \phi} + \frac{\delta\Gamma^{ (n-1) }}{\delta \phi}\frac{\delta\phantom{L}}{\delta L} + \frac{\delta\Gamma^{ (n-1) }}{\delta K}\frac{\delta\phantom{B}}{\delta B} + \frac{\delta\Gamma^{ (n-1) }}{\delta B}\frac{\delta\phantom{K}}{\delta K} \right. + \nonumber                             \\
                                 & +\left.\left(\bar{c}+\bar{\eta}\right)\left(\frac{\delta\phantom{\bar{c}}}{\delta\bar{c}}-\frac{\delta\phantom{\bar{\eta}}}{\delta\bar{\eta}}\right)\right] \;.
\end{align}
Due to the recursive method of the algebraic renormalization technique, results valid for $ \Gamma^{ (1) } $ are equally valid for $ \Gamma^{ (n) } $, and \textit{vice-versa}. Additionally, the linearized Slavnov-Taylor operator~\eqref{eq:linear-st-operator} is nilpotent, and defines a cohomology which is isomorphic to the $ s $-cohomology.

The most general solution of~\eqref{eq:ward-identities} for $n=1$, represents the most general counterterm that $n$-loop radiative corrections can generate. A good way to start is via~\eqref{eq:quantum-st-identity}, due to the nilpotency of the linearized Slavnov-Taylor operators, $ {\mathcal{S}_{\Sigma}}^{ 2 } = 0 $. Its most general solution reads
\begin{equation}
  \label{eq:solution-to-quantum-st}
  \Sigma^{ (1) } = \Delta^0 + \mathcal{S}_{ \Sigma } \Delta^{ -1 }\;.
\end{equation}
$ \Delta^0 $ and $ \Delta^{ -1 } $ are the most general integrated polynomial invariant of the quantum fields and external sources, which are local, power-counting renormalizable, and a 4-form. In particular, $ \Delta^0 $ has ghost number 0, and $ \Delta^{ -1 } $ has ghost number -1. The former belongs to the $ \mathcal{S}_{ \Sigma } $-cohomology, and is given by
\begin{equation}
  \label{eq:delta0}
  \Delta^0 \equiv \int \tr \left( a_1 RR + a_2 RR^* \right) \;,
\end{equation}
where $ a_1 $ and $ a_2 $ are arbitrary renormalization parameters. The latter, also consistent with all the other Ward identities in~\eqref{eq:ward-identities}, is given by
\begin{equation}
  \label{eq:delta-1}
  \Delta^{-1} \equiv \int \tr \left( \alpha_3 Y R + \alpha_4 Y \star R + \alpha_5 Y R^* + \alpha_6 Y \star R^* + \beta \bar{ \chi } R^{\pm} \right) \;,
\end{equation}
where $ \alpha_3 $, $ \alpha_4 $, $ \alpha_5 $, $ \alpha_6 $, and $ \beta $ are also arbitrary renormalization parameters. Finally, the most general counterterm action functional, and solution to~\eqref{eq:ward-identities}, is
\begin{align}
  \label{eq:ct-action}
  \Sigma^{ (1) } & = \int \tr \left\{ \vphantom{^{\pm}} a_1 R R + a_2 R R^* + \alpha_3 \left[ X R + Y \left( D \psi + \left[ c, R \right] \right) \right] + \alpha_4 \left[ X \star R \right. + \right. \nonumber                            \\
                 & \left. + \left. Y \star \left( D \psi + \left[ c, R \right] \right) \right] + \alpha_5 \left[ X R^* + Y {\left( D \psi + \left[ c, R \right] \right)}^* \right] + \alpha_6 \left[ X \star R^* \right. + \right. \nonumber \\
                 & \left. + \left. Y \star {\left( D \psi + \left[ x, R \right] \right)}^* \right] + \beta \left[ BR^{ \pm } + \bar{ \chi } {\left( D \psi + \left[ c, R \right] \right)}^{ \pm }  \right]\right\} \;.
\end{align}

\subsection{Quantum stability}\label{sec:stability;sec:quantum}

To conclude the proof of renormalizability, one has to show the existence of finite parameters $ z_{ \Phi } $, $ z_{ G } $, $ z_{ J } $, such that
\begin{equation}
  \label{eq:multiplicative-renorm}
  \Sigma \left[ \Phi_0, G_0, J_0 \right] = \Sigma \left[ \Phi, G, J \right] + \epsilon\Sigma^{ (1) } \left[ \Phi, G, J \right] \;,
\end{equation}
where
\begin{subequations}%
  \label{eq:field-redef}
  \begin{align}
    \Phi_0 & \equiv z_{ \Phi } \Phi \;\; ; \;\; \Phi \in \left\{ \omega, c, \psi, \phi, \bar{ c }, b, \bar{ \chi }, B, \bar{ \phi }, \bar{ \eta } \right\} \;, \\
    G_0    & \equiv z_{ G } G \;\; ; \;\; G \in \left\{ g_1, \ldots, g_{10} \right\} \;,                                                                       \\
    J_0    & \equiv z_{ J } J \;\; ; \;\; J \in \left\{ \tau, \Omega, E, L, \lambda, K, Z, H, Y, X \right\} \;,
  \end{align}
\end{subequations}
is the multiplicative redefinition of all quantum fields, coupling parameters, and external sources. This requirement guarantees that no infinities are left untamed to all orders in perturbation theory. For our topological symmetry-restored phase of gravity, these $z$-factors are
\begin{subequations}%
  \label{eq:z-factors}
  \begin{align}
    z_{ X }               & = { z_{ H } }^{ -1 } \;,                                                                                                                          \\
    z_{ \omega }          & = z_{ b } = 1 \;,                                                                                                                                 \\
    z_{ g_{ 1 } }         & = 1 + \epsilon { g_{ 1 } }^{ -1 } a_{ 1 } \;,                                                                                                     \\
    z_{ g_{ 2 } }         & = 1 + \epsilon { g_{ 2 } }^{ -1 } a_{ 2 } \;,                                                                                                     \\
    z_{ g_{ 3 } } z_{ X } & = 1 + \epsilon { g_{ 3 } }^{ -1 } \alpha_{ 3 } \;,                                                                                                \\
    z_{ g_{ 4 } } z_{ X } & = 1 + \epsilon { g_{ 4 } }^{ -1 } \alpha_{ 4 } \;,                                                                                                \\
    z_{ g_{ 5 } } z_{ X } & = 1 + \epsilon { g_{ 5 } }^{ -1 } \alpha_{ 5 } \;,                                                                                                \\
    z_{ g_{ 6 } } z_{ X } & = 1 + \epsilon { g_{ 6 } }^{ -1 } \alpha_{ 6 } \;,                                                                                                \\
    z_{ B }               & = 1 + \epsilon \beta = { z_{ K } }^{ -1 } \;,                                                                                                     \\
    z_{ g_{ 7 } }         & = z_{ g_{ 8 } } =  z_{ g_{ 9 } } = z_{ g_{ 10 } }  = { z_{ H } }^{ 2 } \;,                                                                        \\
    z_{ \bar{ \phi } }    & = z_{ \tau } = z_{ L } = {z_{ \bar{ c } }}^{ 2 } = z_{ \lambda } z_{ \bar{ \chi } } = z_{ Y } z_{ Z } = {z_{ \phi } }^{ -1 } \;,                  \\
    z_{ \bar{ c } }       & = z_{ \Omega } = z_{ \bar{ \eta } } = z_{ B } z_{ \lambda } = z_{ Y } z_{ H } = z_{ E } z_{ \phi } = {z_{ \psi } }^{ -1 } = {z_{ c } }^{ -1 } \;.
  \end{align}
\end{subequations}
Clearly, the gauge field $ \omega $ does not renormalize. This is a feature of TYM due to the lack of local field equations. The physical coupling, $ g_1 $ and $ g_2 $, renormalize with the $ s $-cohomology, while the non-physical ones, $ g_3, \ldots, g_{10} $, renormalize with $ s $-boundaries. This is exactly what is to be expected.

\section{Conclusions}%
\label{sec:conclusions}

In this work, we proposed an $ SO \left( 4 \right) $ TYM theory, defined by~\eqref{eq:tg+sb}, as topological phase of gravity. The TYM BRST symmetry transformations~\eqref{eq:top_grav_brst}, play a pivotal role in this context. They \textquote{freeze} the local gravitational dynamics into the physically trivial $s$-boundary term~\eqref{eq:s-boundary_action}. In this topological phase, the gravitational observables that remain are related to the smoothness of spacetime --- which automatically makes them conformally invariant.

The unprecedented $ s $-boundary term~\eqref{eq:s-boundary_action}, vital to the connection to gravity, has the potential to spoil the renormalizability features of $ SO(4) $ TYM.\@ In Section~\ref{sec:quantum}, the quantum stability of the proposed model is worked out in the extended (A){}SDL gauge~\eqref{eq:extended-asdlg}. Remarkably, the model remains renormalizable to all orders in perturbative theory, and contains 7 independent renormalizations~\eqref{eq:ct-action}.

Key features of topological quantum field theories remain explicit in the present model. For instance, the complete set of Ward identities of traditional QTYM, in the (A){}SDL gauge, is present here --- albeit corrected to account for the presence of~\eqref{eq:s-boundary_action}. We found that this strong set of symmetries implies that the gauge field does not renormalize, $ z_{ \omega } = 1 $. This result is compatible with the vanishing of $ \langle \omega(x) \omega(y) \rangle $ to all orders in perturbation theory. The latter is a known feature of traditional QTYM in the (A){}SDL gauge~\cite{sadovski2017c,sadovski2018a}. This is yet another reflection of the physical content of the theory being non-local in the bulk and/or living in the boundary.

The connection between the proposed topological phase and the Lovelock-Cartan family of gravity theories is explained in Section~\ref{ssec:from_top_to_grav;sec:top_grav}. And, it boils down to a dynamical implementation of the horizontal condition, $ \psi = \phi = 0 $. This condition deforms the TYM BRST into the traditional YM BRST.\@ Local observables in the bulk --- incompatible with the TYM BRST --- are now allowed by the YM BRST.\@ Most remarkably, the proposed implementation also identifies these bulky local degrees of freedom as gravitational ones.

The detailed nature of the mass parameter $ \mu $ in~\eqref{eq:physicalsource2} is left open. Several mass generation mechanisms, well known in the context of non-Abelian gauge theories, can be employed to explain its origin, and to predict its current theoretical value. The viability of the model can then be tested by comparison to most current the experimental value of $ m_{ \text{P} } $, and the most current observational value of $ \Lambda^2 $, via~\eqref{eq:correspondence}. This investigation will be carried out by the authors in a follow-up work.

\section*{Acknowledgements}
Author R.~F.~Sobreiro would like to acknowledge the Coordena{\c{c}}{\~{a}}o de Aperfei{\c{c}}oamento de Pessoal de N{\'{\i}}vel Superior - Brasil
(CAPES) - Finance Code 001, for financial support.

\printbibliography{}

\end{document}